\documentclass[aps,amsfonts,amsmath,prd,preprint,nofootinbib]{revtex4}
\pdfoutput=1
\newcommand{\beq}{\begin{equation}}
\newcommand{\eeq}{\end{equation}}
\usepackage{graphicx}
\usepackage{natbib}
\usepackage[utf8]{inputenc}
\begin{document}

\title{Gravitational wave background from mergers of large primordial black holes}
\author{Heling Deng}
\email{heling.deng@asu.edu}
\affiliation{Physics Department, Arizona State University, Tempe, AZ 85287, USA}

\begin{abstract}
The Peters formula, which tells how the coalescence time of a binary
system emitting gravitational radiation is determined by the initial size
and shape of the elliptic orbit, is often used
in estimating the merger rate of primordial black holes and the gravitational wave background from the mergers. Valid as it is in some interesting
scenarios, such as the analysis of the LIGO-Virgo events,
the Peters formula fails to describe the coalescence time if the orbital period of the binary exceeds
the value given by the formula. This could underestimate the event rate
of mergers that occur before the cosmic time $t\sim 10^{13}\ \text{s}$.
As a result, the energy density spectrum of the gravitational wave background could develop a peak, which is from mergers occurring at either $t\sim 10^{13}\ \text{s}$ (for black holes with mass $M\gtrsim 10^8 M_\odot$) or $t\sim 10^{26}(M/M_\odot)^{-5/3}\ \text{s}$ (for $10^5 M_\odot \lesssim M\lesssim 10^8 M_\odot$). This
can be used to constrain the fraction of dark matter in
primordial black holes (denoted by $f$) if potential probes
(such as SKA and U-DECIGO) do not discover such a background,
with the result $f\lesssim 10^{-6}\text{-}10^{-4}$ for the mass range $10\text{-} 10^9M_\odot$. We then consider the effect of mass accretion onto primordial black holes at redshift $z\sim 10$, and find that the merger rate could drop significantly at low redshifts. The spectrum of the gravitational wave background thus gets suppressed at the high-frequency end. This feature might be captured by future detectors such as ET and CE for initial mass $M= \mathcal{O}(10\text{-}100) M_\odot$ with $f\gtrsim 10^{-4}$.

\end{abstract}

\maketitle
\section{Introduction}

Primordial black holes (PBHs) are hypothetical black holes formed
in the early universe before any large scale structures and galaxies.
Unlike astrophysical black holes formed by dying stars at low redshifts,
which are expected to have initial mass slightly larger than the solar
mass ($M_{\odot}\sim10^{33}$ g), PBHs can in principle be born with
any mass beyond the Planck mass ($M_{\rm Pl}\sim10^{5}$ g) during the
radiation era. PBHs have drawn considerable attention in recent years, mostly
because the LIGO-Virgo Collaboration has so far detected around 50
signals, most of which are believed to be gravitational waves emitted
from inspiraling and merging black holes of mass $\mathcal{O}(10\text{-}100)M_\odot$ \cite{LIGOScientific:2018mvr,LIGOScientific:2020ibl}. The origin of these black
holes is so far unknown. Particularly, some of these black holes lie within the ``pair instability mass gap", which
demands further explanation if they were formed by stellar collapse \cite{Woosley:2016hmi,Belczynski:2016jno,Spera:2017fyx,Giacobbo:2017qhh}. A fascinating speculation is that LIGO-Virgo has detected PBHs \cite{Bird:2016dcv,Sasaki:2016jop,Clesse:2016vqa}.\footnote{Discussion of distinguishing PBHs from astrophysical black holes by future detectors can be found in, e.g., refs \cite{Chen:2019irf,Mukherjee:2021ags}.}

The
physical picture of the formation of PBH binaries in the early universe
is rather simple \cite{Sasaki:2016jop}. If PBHs are distributed randomly in space,
they rarely interact with each other, with their number density
simply diluted by the cosmic expansion. However, with some small probability,
two PBHs can form so close to each other that their gravitational
attraction defeats the Hubble stretch at some point during the radiation
era. Due to the surrounding disturbance, the two black holes can evade
head-on collision and form a binary with an elliptic orbit.
The major axis of the ellipse decreases due to energy loss by the emission of
gravitational radiation, and the two black holes will thus merge into
one eventually. Gravitational radiation from some mergers might
have been detected by LIGO-Virgo if the signals were sufficiently strong, if the
ringdown happened to be at the frequency band of the detectors, and if
the coalescence times of the binaries were around the age of the
universe. Given the PBH mass and abundance, one is able to estimate the number of signals that could have been caught by the detectors.

It is suggested by the merger rate inferred by LIGO-Virgo that if the detected black holes were all or partially primordial, they are expected to constitute
around 0.1\% of the dark matter \cite{Sasaki:2016jop,Raidal:2017mfl,Ali-Haimoud:2017rtz,Vaskonen:2019jpv,Garriga:2019vqu,Hutsi:2020sol,Deng:2021ezy,Franciolini:2021tla}. With this abundance, mergers of binary PBHs over the cosmic history should have generated a gravitational wave background (GWB) within the reach of the Laser Interferometer
Space Antenna (LISA) \cite{LISA:2017pwj} and the future runs of LIGO \cite{LIGOScientific:2016fpe}. The non-detection
of this background can thus rule out PBHs as an explanation of the
merger events detected so far \cite{Wang:2016ana,Raidal:2017mfl}. In a similar manner, ref. \cite{Wang:2019kaf} further studies
the mergers of PBHs within the mass range $10^{-8}\text{-}1 M_{\odot},$
with the result that the non-detection of GWB by the Big Bang Observer
(BBO) \cite{Harry:2006fi} can significantly improve the current limit on the PBH abundance
for the mass range $10^{-6}\text{-}1 M_{\odot}$.

In the present work, we revisit the merger rate of large PBHs with mass $M\gtrsim10M_{\odot}$. This
was motivated by noticing that the often adopted Peters formula \cite{Peters:1964zz}
fails to account for the coalescence time of a binary in some regime. The formula,
which tells how the coalescence time is determined by the initial
size and shape of the elliptic orbit, is often used in
calculating the merger rate of PBHs and GWB from the mergers. However, it underestimates the coalescence time when the orbital period (or the free-fall time) of
the binary is larger than the value given by the formula, which, roughly speaking, is the case for mergers that take place before redshift $z\sim1000$ if PBHs constitute a tiny fraction
of the dark matter.

As we will see, merger events that occur this early could significantly affect the energy density spectrum
of the resulting GWB.
It is the task of this paper to find the relevant spectra, and constrain
the PBH abundance for $M\gtrsim10M_{\odot}$
if no GWB is detected in future missions. We will also consider a toy model of mass accretion on PBHs and discuss how the GWB spectrum would be affected in some special cases, which could possibly be captured by future detectors.

The rest of the paper is organized as follows. In section \ref{II} we estimate the merger rate of PBH binaries, considering the regime where the Peters formula fails. In section \ref{III} we discuss how this affects the spectrum of GWB from PBH mergers and how it can be used to constrain the PBH abundance if future missions do not see this background. The effect of mass accretion will be considered in section \ref{IV}. Conclusions are summarized and discussed in section
\ref{V}. We set $c=G=1$ throughout the paper.

\section{PBH merger rate  \label{II}}

If we assume that PBHs are distributed randomly in space in the early universe, they generally do not run across each other but simply follow the Hubble flow. However, if two PBHs happen to be separated by a sufficiently short distance, they could decouple from the Hubble flow and move towards each other due to the gravitational attraction. 
This typically occurs when the free-fall time of the two black holes is surpassed by the ever-increasing Hubble time, when Newtonian physics comes into play \cite{Sasaki:2018dmp}.  After then, due to the disturbance of the environment, a typical example being a third neighboring black hole exerting a tidal
torque on them, the two black holes
can evade head-on collision and form a binary inspiraling in a highly eccentric elliptic
orbit \cite{Nakamura:1997sm,Ioka:1998nz,Sasaki:2016jop}. The cumulative effect of all other surrounding PBHs was studied in, e.g., refs. \cite{Ali-Haimoud:2017rtz,Raidal:2018bbj,Kocsis:2017yty}. As pointed out in ref. \cite{Garriga:2019vqu}, although considering the nearest PBH only does not provide a good description of the probability distribution of the orbital eccentricity when it is close to 1, the torque is indeed dominated by this PBH, and the merger rate of the binary is not significantly affected by the inclusion of other neighbors.\footnote{Strictly speaking, this is true when PBHs merely constitute a tiny part of the dark matter, which is the case we are interested in in the present work. If the PBH abundance is sufficiently large, the binary could easily be disrupted by the closest PBH, and all surrounding PBHs become relevant. Furthermore, the merger rate would also be suppressed due to the formation of PBH clusters. More details can be found in refs. \cite{Raidal:2018bbj,Vaskonen:2019jpv}. I would like to thank Hardi Veerm\"ae and Ville Vaskonen for pointing this out.}

In addition to the surrounding black holes, cosmological density perturbations on scales larger than the binary at its formation should also generate a tidal torque. This would certainly affect the eccentricity of the binary's orbit and thus the merger rate. Although the magnitude of perturbations on scales smaller than those observed in CMB is unknown, it might be extrapolated from the CMB results \cite{Ali-Haimoud:2017rtz}. It was found that the effect from perturbations would be crucial if the PBH abundance is small (compared to the rescaled variance of the perturbations). 

For simplicity, we shall assume that all PBHs have the same mass denoted
by $M$. Let $f$ be the fraction of dark
matter in PBHs. The physical number density of PBHs at dust-radiation
equality (redshift $z_{\rm eq}\approx3000$) is
\begin{equation}
n\sim\frac{f\rho}{M}\sim\bar{x}^{-3},\label{eq:n}
\end{equation}
where $\rho$ is the dark matter density at $z_{\rm eq}$, and $\bar{x}$
is the average physical distance between two PBHs at $z_{\rm eq}$. Let $x/z_{\rm eq}$ be the comoving distance between two random neighboring PBHs when they are formed in the very early universe. When the two black holes decouple from the Hubble flow, their physical separation is approximately the semi-major axis of the elliptic orbit, and can be estimated as \cite{Nakamura:1997sm}
\begin{equation}
a\approx\frac{x^{4}}{f\bar{x}^{3}}.\label{eq:a}
\end{equation}

As two PBHs run in the orbit, some energy is carried away by the emission of gravitational
radiation. As this occurs, the value of the semi-major axis of the orbit
decreases, and the two black holes will eventually merge into one.
In the literature, the coalescence time of a PBH binary is often taken to
be the result obtained in ref. \cite{Peters:1964zz} by Peters (which will be referred to
as the Peters formula hereafter):
\begin{equation}
t_{\rm P}=\frac{3}{170}\frac{a^{4}}{M^{3}}\left(1-e^{2}\right)^{7/2},\label{eq:merge1}
\end{equation}
where $e$ is the initial eccentricity of the
orbit. This
formula is obtained by assuming that the coalescence time is (much)
larger than the period of the orbit. Under this assumption, quantities such as the rate of 
energy loss during the coalescence can be found in a closed form
by averaging over a period. Note that $t_{\rm P}\to 0$ when $e\to 1$, whereas the orbital period is independent of the eccentricity. This implies there exists a regime where the assumption behind the Peters formula would fail. More precisely, since the orbital period, which is comparable to the free-fall time of the two black holes, is roughly
given by
\begin{equation}
t_{\rm ff}=\left(\frac{a^{3}}{M}\right)^{1/2},\label{eq:merge 2}
\end{equation}
the coalescence time can be estimated by the Peters formula only if $t_{\rm P}>t_{\rm ff}$, otherwise
the two black holes simply fall into each other without finishing
a complete period, and the coalescence time is thus dominated by  $t_{\rm ff}$. Therefore, two black holes in a binary should merger at
\begin{equation}
t\sim t_{\rm P}+t_{\rm ff}.\label{eq:t_new}
\end{equation}

Now let $y$ be the physical distance from the third nearby
PBH to the binary at $z_{\rm eq}$. By refs. \cite{Sasaki:2016jop, Atal:2020yic}, the probability density that a random PBH belongs to a binary and the merger occurs at cosmic time $t$ can be estimated as
\begin{equation}
P(t) = \mathcal{O}(10)n^{2}\int_{y_{\rm min}}^{y_{\rm max}} x^{2}y^{2}\left|\frac{\text{d}x}{\text{d}t}\right|\text{d}y,\label{eq:P}
\end{equation}  
where $t$ should also be regarded as the coalescence time. In the following two subsections, we will describe how the initial eccentricity, and thus the merger probability, of a binary are determined by the third nearby black hole and by cosmological density perturbations.

\subsection{Effect from the third PBH}
   
The tidal torque from the third nearby black hole dislocates the two PBHs in the binary, leading to a semi-minor axis $b\approx (x/y)^3a$ \cite{Nakamura:1997sm}. Then the eccentricity of the binary at its formation can be estimated as \cite{Nakamura:1997sm,Ioka:1998nz}
\begin{equation}
e\approx\sqrt{1-\left(\frac{x}{y}\right)^{6}}.\label{eq:e}
\end{equation}

\begin{figure}
\includegraphics[scale=0.5]{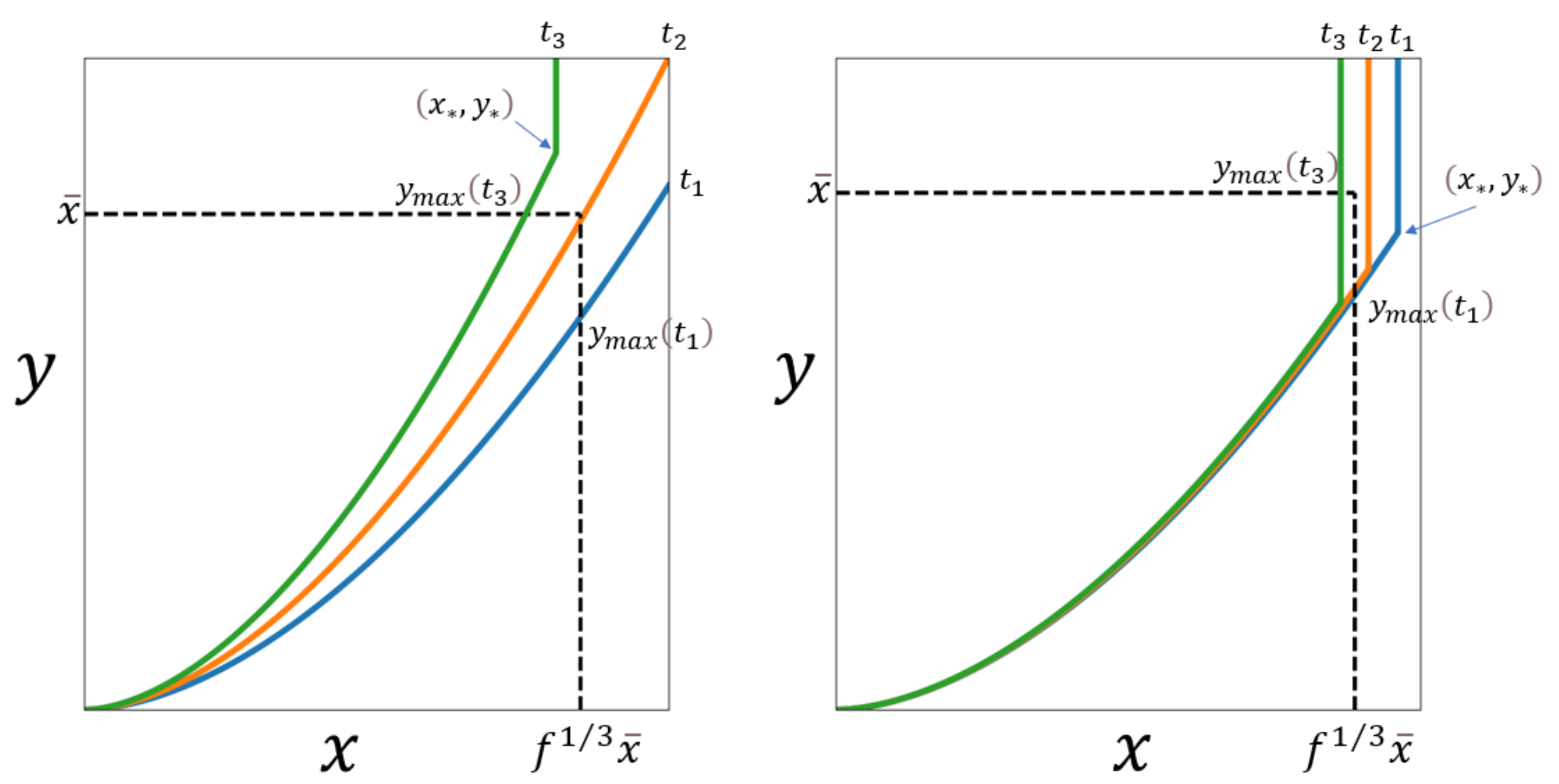}

\caption{\label{fig:xy}Several examples showing how the piecewise function
(\ref{eq:x}) intersects the boundaries $x=f^{1/3}\bar{x}$ and $y=\bar{x}$.
In both plots we have $t_{1}>t_{2}>t_{3}$. $Left$: Curves with fixed
coalescence times $t_{1}$ (blue) and $t_{2}$ (orange) are in case
(a); $t_{3}$ (green) curve is in case (b). Since the transition points $(x_{\ast},y_{\ast})$ of the three curves
are outside the bounded region (here we only show $(x_{\ast},y_{\ast})$
for $t_{3}$), the coalescence times in these three examples are dominated
by the Peters formula. $Right$: $t_{1}$
(blue) and $t_{2}$ (orange) curves are in case (a); $t_{3}$
(green) curve is in case (c). The maximum value of $y$ increases from $y_{\rm max}(t_1)(<\bar{x})$ to $y_{\rm max}(t_3)(=\bar{x})$ as the transition point $(x_{\ast},y_{\ast})$ enters the bounded region. $t_1$ and $t_2$ are dominated by the Peters formula, whereas $t_3$ is dominated by the
free-fall time $t_{\rm ff}$. }
\end{figure}

In order to evaluate the integral (\ref{eq:P}), we first express $x$ as a function
of $y$ and $t$. By eqs. (\ref{eq:n})-(\ref{eq:t_new}) and (\ref{eq:e}), the coalescence time given $M,x$ and $y$ is
\begin{equation}
t\sim\frac{3\rho^{4}}{170M^{7}}\frac{x^{37}}{y^{21}}+\frac{\rho^{3/2}}{M^{2}}x^{6}.\label{eq:t-1}
\end{equation}
Then for a fixed $t$, we have
\begin{equation}
y=\left(\frac{3\rho^{5/2}}{170M^{5}}\frac{x^{37}}{M^{2}\rho^{-3/2}t-x^{6}}\right)^{1/21},\label{eq:y}
\end{equation}
which goes to infinity when 
\begin{equation}
x\to \frac{M^{1/3}t^{1/6}}{\rho^{1/4}}\equiv x_{\ast}. \label{xstar}
\end{equation}
We can then approximate (\ref{eq:y}) by the following piecewise
function
\begin{equation}
x/x_{\ast}\sim\begin{cases}
\left(y/y_{\ast}\right)^{21/37}, & y<y_{\ast},\\
1, & y>y_{\ast},
\end{cases}\label{eq:x}
\end{equation}
where $y_{\ast}\equiv\left(3M^{16/3}\rho^{-21/4}t^{31/6}/170\right)^{1/21}$.
Several examples of this function are shown in fig. \ref{fig:xy}.

The integral in eq. (\ref{eq:P}) is bounded by three conditions: $x<f^{1/3}\bar{x}$, $y<\bar{x}$ and $x<y$. The first one is imposed to make sure that the binary
is formed during the radiation era, otherwise the two PBHs are unable
to decouple from the Hubble flow because the free-fall time $t_{\rm ff}$ is proportional to the Hubble time during the dust era. The second bound reflects
the fact that, if PBHs are randomly distributed in space, the probability
that two PBHs are separated by a distance larger than $\bar{x}$ is
exponentially suppressed. The third condition is required in order for the third black hole not to be part of the binary. In evaluating the integral, the third condition gives the minimum value of $y$ and does not contribute much in scenarios we are interested in. The merger probability $P(t)$ is mainly determined
by the maximum value of $y$, which in principle is determined by
the intersection of eq. (\ref{eq:y}) and the two boundaries $x=f^{1/3}\bar{x}$
and $y=\bar{x}$. 

As can be seen from fig. \ref{fig:xy},
the simplified piecewise function (\ref{eq:x}) intersects the two boundaries
in three ways: (a) $x=\left(y/y_{\ast}\right)^{21/37}x_{\ast}$ intersects
$x=f^{1/3}\bar{x}$ (blue and orange curves in both panels); (b) $x=\left(y/y_{\ast}\right)^{21/37}x_{\ast}$
intersects $y=\bar{x}$ (green curve in the left panel); (c) $x=x_{\ast}$ intersects $y=\bar{x}$ (green curve in the right panel). By finding
out $y_{\rm max}$ in these three
cases, the PBH merger probability can be estimated as
\begin{equation}
P(t)\approx \frac{0.05f}{t}\begin{cases} 
\left(\frac{t}{t_c}\right)^{-1/7}, & \text{case (a)},\\
\left(\frac{t}{t_c}\right)^{3/37}, & \text{case (b)},\\
\left(\frac{t}{10^{11}\ \text{s}}\right)^{1/2}, & \text{case (c)},
\end{cases}\label{eq:P-1}
\end{equation}
where $t_c= 0.02 f^{7}(\rho^4 M^5)^{-1/3}\sim 10^{42}f^{7}(M/M_\odot)^{-5/3}\ \text{s}$ is the transition time from case (a) to case (b).

Were it one of the first two cases, the coalescence time is dominated by
the Peters formula. This is the case for, e.g., merger events that occur near the present epoch. If, however, as we decrease the value of $t$ (from, say, the present time) and find that case (a) is directly transitioned to (c), the maximum value of $y$ suddenly increases from
$y_{\rm max}<\bar{x}$ to $y_{\rm max}=\bar{x}$ (see the right panel in fig.
\ref{fig:xy}), which leads to a sudden increase in $P(t)$. This occurs
when $x_*$ becomes smaller than $f^{1/3}\bar{x}$, which gives $t<\rho^{-1/2}\sim10^{13}\ \text{s}$.
In other words, if $y_{\ast}<\bar{x}$ at $t\sim10^{13}\ \text{s}$,
the merger probability can increase significantly at $t\lesssim10^{13}\ \text{s}$ compared to the case without the term $t_{\rm ff}$ included in the coalescence time.

Physically, this means PBH binaries could have a significantly larger
merger rate before the time of recombination than previously expected. As discussed before, a PBH binary is formed at a cosmic time comparable the free-fall time $t_{\rm ff}$. Hence if the coalescence time is dominated by $t_{\rm ff}$, two black holes would merge in around a Hubble time after the formation of the binary. Since the last binaries should be formed at around dust-radiation equality, the last mergers that are affected by the ``new" term $t_{\rm ff}$ in eq. (\ref{eq:t_new}) should take place at redshift $z\sim 1000$. For merger events after this time, the merger probability is given by either case (a) or case (b).

If the coalescence time is given by the Peters formula, then as we decrease the value of $t$, case (a) is transitioned to case (b) at $t_{c}$. If this is earlier than $\sim 10^{13}\ \text{s}$, then case (a) would directly turn into case (c): $t_{c} \lesssim 10^{13}\ \text{s}$ gives
\begin{equation}
f^7\left(\frac{M}{M_\odot}\right)^{-5/3}\lesssim 10^{-28}.\label{condition}
\end{equation}
Only when this condition is satisfied can there be an abrupt increase in merger rate at $\sim 10^{13}\ \text{s}$ as we go back in time. 
 
\subsection{Effect from cosmological perturbations}
As mentioned earlier, the additional tidal torque from cosmological density perturbations on PBH binaries would be important if PBHs constitute a tiny part of the dark matter. This is the case if we are interested in constraining the PBH abundance with potential gravitational wave (GW) detectors at different frequency ranges. Following ref. \cite{Ali-Haimoud:2017rtz}, we let $\sigma=0.005$ be the rescaled variance of density perturbations on a scale larger than the binary when it decouples from the Hubble flow. The binary acquires an extra angular momentum, and its eccentricity can typically be approximated by \cite{Ali-Haimoud:2017rtz}
\begin{equation}
e\approx\sqrt{1-\left(\frac{x}{y}\right)^{6}-\left(\frac{\sigma}{f}\right)^2\left(\frac{x}{\bar{x}}\right)^{6}}.
\end{equation}
Thus the coalescence time of the binary becomes
\begin{equation}
t\sim\frac{3\rho^{4}}{170M^{7}}\frac{x^{37}}{y^{21}}\left[1+\left(\frac{\rho\sigma}{M}\right)^2y^{6}\right]^{7/2}+\frac{\rho^{3/2}}{M^{2}}x^{6}.\label{eq:t-2}
\end{equation}
If it is dominated by the Peters formula, we have
\begin{equation}
y=\left[\left(\frac{170M^7}{3\rho^4}\frac{t}{x^{37}}\right)^{2/7}-\left(\frac{\rho\sigma}{M}\right)^2\right]^{-1/6},\label{eq:y1}
\end{equation}
which goes to infinity when
\begin{equation}
x\to \left(\frac{170M^{14}}{3\sigma^7 \rho^{11}}t\right)^{1/37}\equiv x_*^{(1)}.
\end{equation}
If the coalescence time is dominated by the free-fall time, then $y$ goes to infinity at
\begin{equation}
x\to \frac{M^{1/3}t^{1/6}}{\rho^{1/4}}\equiv x_\ast^{(2)}.
\end{equation}
Therefore, similar to eq. (\ref{eq:x}), $x$ as a function of $y$ with $t$ fixed can be approximated by the following piecewise function:
\begin{equation}
x/x_{\ast}\sim\begin{cases}
\left(y/y_{\ast}\right)^{21/37}, & y<y_{\ast},\\
1, & y>y_{\ast},
\end{cases}\label{eq:x1}
\end{equation}
where $x_\ast=\text{min}(x_\ast^{(1)},x_\ast^{(2)})$, and $y_\ast$ is now obtained by inserting $x=x_\ast$ in eq. (\ref{eq:y1}).

The merger probability can then be estimated in a form similar to eq. (\ref{eq:P-1}), with the result
\begin{equation}
P(t)\approx \frac{0.05f}{t}\begin{cases} 
\left[1+\left(\frac{\sigma}{f}\right)^2\left(1-\left(\frac{t}{t_c^\prime}\right)^{-2/7}\right)\right]^{-29/37}\left(\frac{t}{t_c^\prime}\right)^{-1/7}, & \text{case (a)},\\
\left(\frac{t}{t_c^\prime}\right)^{3/37}, & \text{case (b)},\\
\frac{60\rho}{M}x_\ast^3\left(37-31\frac{\rho^{3/2}x_\ast^6}{M^2t}\right)^{-1}. & \text{case (c)},
\end{cases}\label{eq:P-4}
\end{equation}
where $t_c^\prime = \left[1+\left(\sigma/f\right)^2\right]^{7/2}t_c$ is the transition time from case (a) to case (b). If $f\ll\sigma$, $t_c^\prime\approx 10^{42}\sigma ^7(M/M_\odot)^{-5/3}\ \text{s}$. If $f\gg\sigma$, $t_c^\prime\approx t_c$, and $P(t)$ is reduced to eq. (\ref{eq:P-1}). This means the effect from density perturbations is only relevant when the PBH abundance is small. In conclusion, to estimate the PBH merger probability $P(t)$ given $M$ and $f$, we should first compare the values of $x_\ast^{(1)}$ and $x_\ast^{(2)}$, then figure out how the piecewise function (\ref{eq:x1}) intersects the bounds $x=f^{1/3}\bar{x}$ and $y=\bar{x}$, and then find $P(t)$ by eq. (\ref{eq:P-4}).

As before, if we decrease the value of $t$ and find the transition point $(x_\ast,y_\ast)$ entering the bounded region from $x=f^{1/3}\bar{x}$, the merger rate typically acquires a sudden increase. The time when this occurs is determined by $x_\ast^{(1)}\sim f^{1/3}\bar{x}$ or $x_\ast^{(2)}\sim f^{1/3}\bar{x}$, whichever gives an earlier time. The latter relation leads to our result from the last subsection: $t\sim 10^{13}\ \text{s}$, and the former gives $t\sim t_c^\prime$.  As we will see in the next section, if PBHs constitute merely a minor part of the dark matter, the energy density spectrum of the
gravitational wave background (GWB) from PBH mergers for $M\gtrsim 10^{5} M_\odot$ could develop an extra peak, which is from mergers that occur at $t_\ast \sim \text{min}(t_c^\prime, 10^{13}\ \text{s})$.   

\section{GWB from PBH mergers  \label{III}}

We will now study the GWB energy density spectrum from the mergers of large PBHs. By ref. \cite{Phinney:2001di},
the GWB spectrum from events emitting gravitational radiation is in general given by
\begin{equation}
\Omega_{\rm GW}(\nu_{\rm d})=\frac{\nu_{\rm d}}{\rho_{c}}\int_0^{z_{\rm max}} N(z)\left.\frac{\text{d}E_{\rm GW}(\nu_{\rm s})}{\text{d}\nu_{\rm s}}\right|_{\nu_{\rm s}=\nu_{\rm d}(1+z)}\text{d}z,\label{eq:Omega}
\end{equation}
where $N(z)\text{d}z$ is the comoving number density of events
that occur within the redshift interval $(z,z+\text{d}z)$, $\rho_{c}$ is the critical
density of the universe, $\nu_{\rm d}$ and $\nu_{\rm s}$ are 
the GW frequencies in the detector frame and the source frame, respectively, $\text{d}E_{\rm GW}/\text{d}\nu_{\rm s}$ is the GW energy spectrum from a single binary black hole merger, and $z_{\rm max}\equiv \nu_3/\nu_{\rm d}-1$ with $\nu_3$ defined in eq. (\ref{dEdnu2}).
For PBH mergers from the mechanism described above, $N(z)$ can be evaluated by
\begin{equation}
N(z)=n z_{\rm eq}^{-3} P(t)\frac{\text{d}t}{\text{d}z},
\end{equation}
where $P(t)$ is given by eq. (\ref{eq:P-4}) (note that $n$ is the physical number density of PBHs at $z_{\rm eq}$). Here the redshift $z$
and the cosmic time $t$ is related by $\text{d}t/\text{d}z=[(1+z)H(z)]^{-1}$,
where $H(z)$ is the Hubble parameter at $z$. The inspiral-merger-ringdown energy spectrum is given by \cite{Ajith:2007kx}
\begin{equation}
\frac{\text{d}E_{\rm GW}(\nu_{\rm s})}{\text{d}\nu_{\rm s}}=\frac{2^{-1/3}\pi^{2/3}M^{5/3}}{3}\begin{cases}
\nu_{\rm s}^{-1/3}, & \nu_{\rm s}<\nu_{1},\\
\nu_{1}^{-1}\nu_{\rm s}^{2/3}, & \nu_{1}<\nu_{\rm s}<\nu_{2},\\
\nu_{1}^{-1}\nu_{2}^{-4/3}\nu_{\rm s}^{2}\left[1+4\left(\frac{\nu_{\rm s}-\nu_{2}}{\sigma}\right)^{2}\right]^{-2}, & \nu_{2}<\nu_{\rm s}<\nu_{3},\\
0, & \nu_{3}<\nu_{\rm s},
\end{cases} \label{dEdnu}
\end{equation}
where frequencies
$\nu_{1},\nu_{2},\nu_{3}$ and $\sigma$ are of the form
\begin{equation}
2\pi M\nu_{i}=\frac{a_{i}}{16}+\frac{b_{i}}{4}+c_{j},\label{dEdnu2}
\end{equation}
with the coefficients $a_{i},b_{i}$ and $c_{i}$ given in table I of ref. \cite{Ajith:2007kx}. Specifically, we have $\nu_1\approx 0.02M^{-1}$, $\nu_2\approx 0.04M^{-1}$ and $\nu_3\approx 0.06M^{-1}$. In the unit of Hz, $M^{-1}\approx 2\times 10^5 (M/M_\odot)^{-1}\ \text{Hz}$.

If the free-fall time and cosmological perturbations are not taken into consideration, the GWB spectrum from PBH mergers is dominated by mergers that occur near the present time and has a peak near the frequency $\nu_2 
\sim 10^4 (M/M_\odot)^{-1}\ \text{Hz}$. Due to the increase of merger rate when case (a) is transitioned to case (c) in eq. (\ref{eq:P-4}), another peak develops
at $\sim z_{\ast}^{-1}\nu_2$, which is from mergers at redshift $z_\ast$ corresponding to the cosmic time $t_\ast \sim \text{min}(t_c^\prime, 10^{13}\ \text{s})$.  The non-detection of this background
would then impose constraints on the density of PBHs. 

\begin{figure}
\includegraphics[scale=0.33]{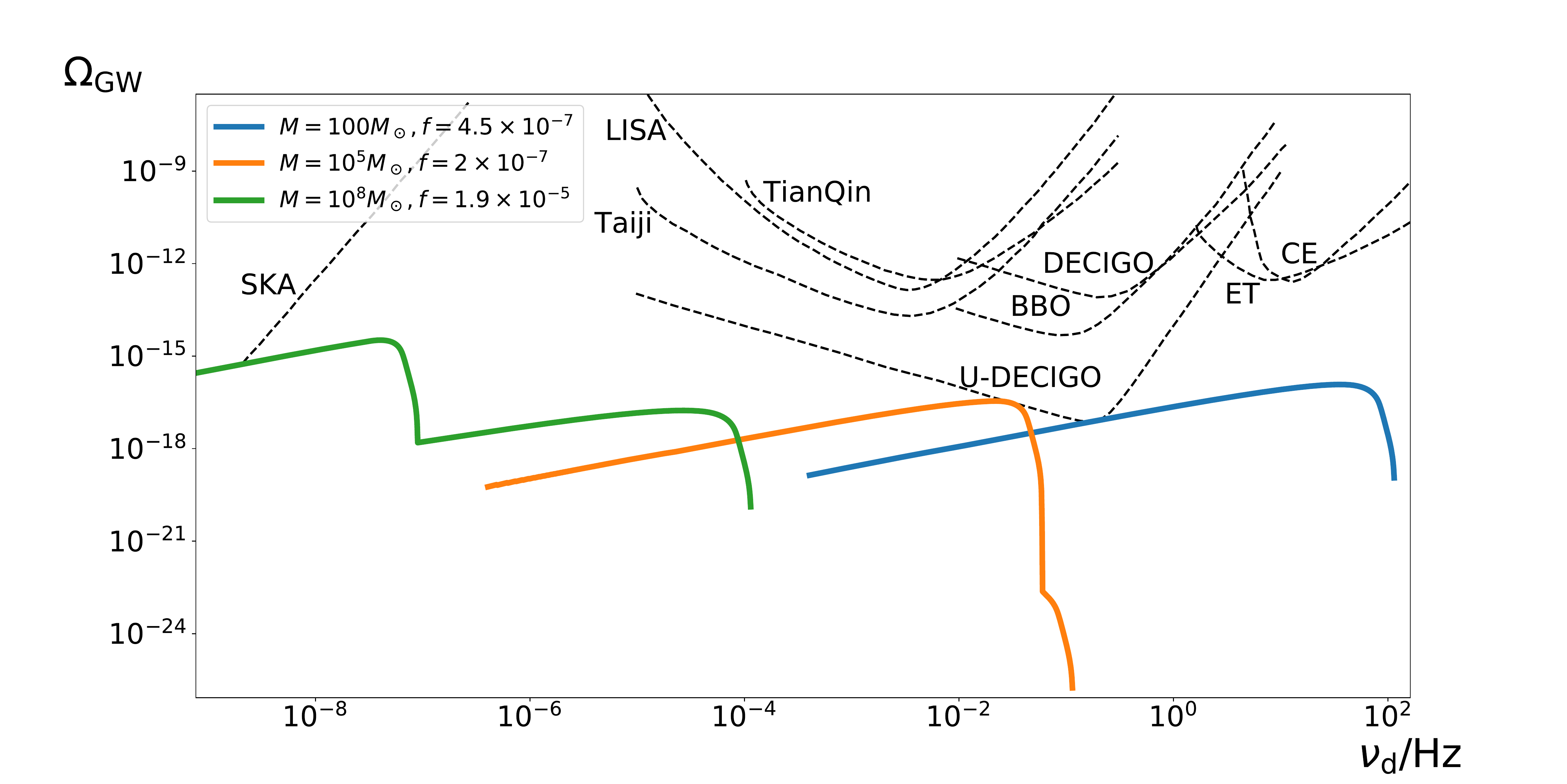}

\caption{\label{fig:Omega} The three colored solid curves are examples of the energy
density spectrum of GWB from mergers of PBH binaries. Each spectrum
is determined by the PBH mass ($M$) and the fraction of dark matter
in PBHs ($f$), both specified in the plot. The dashed curves are sensitivity curves of potential GW projects SKA, TianQin, LISA, Taiji, DECIGO, BBO, U-DECIGO, CE and ET. Parameters $M$ and $f$ are specifically chosen
such that the corresponding signals are just out of the reach of the
detectors. The peak at $\sim 100\ \text{Hz}$ in the blue spectrum and the peak  at $\sim 10^{-4}\ \text{Hz}$ in the green spectrum are from mergers near the present time, whereas other two peaks in the orange and the green spectra are from mergers near the cosmic time $t_\ast \sim \text{min}(t_c^\prime, 10^{13}\ \text{s})$.}
\end{figure}

In fig. \ref{fig:Omega} we plot the designed sensitivity curves of some potential GW probes: the Square Kilometer Array (SKA) \cite{Smits:2008cf, Janssen:2014dka}, TianQin \cite{TianQin:2015yph,Liang:2021bde}, LISA, Taiji \cite{Hu:2017mde}, the DECi-hertz Interferometer Gravitational wave Observatory (DECIGO) \cite{Seto:2001qf,Kawamura:2011zz}, BBO, U-DECIGO \cite{Kudoh:2005as}, the Einstein Telescope (ET) \cite{punturo2010einstein,Maggiore:2019uih} and the Cosmic Explorer (CE) \cite{LIGOScientific:2016wof} (dashed curves). We also show
three examples of $\Omega_{\rm GW}$ (solid curves). The PBH abundances in these examples
are chosen such that the spectra barely intersect
the sensitivity curves, which means each possible GWB is just beyond the
reach of the designed detectors.

The behavior of the three examples in fig. \ref{fig:Omega} can be understood as follows. If $f\ll\sigma\approx 0.005$, then by definition,  $t_c^\prime \sim 10^{26}(M/M_\odot)^{-5/3}\ \text{s}$. If $t_c^\prime \lesssim 10^{13}\ \text{s}$, which gives $M\gtrsim 10^8 M_\odot$, the PBH merger rate could increase at cosmic time $t\sim 10^{13}\ \text{s}$ compared to the case without the free-fall time $t_{\rm ff}$ included in the coalescence time. If $t_c^\prime$ is earlier than the present time, i.e., $10^{13}\ \text{s} \lesssim t_c^\prime \lesssim 4\times 10^{17}\ \text{s}$, which gives $10^5 M_\odot \lesssim M \lesssim 10^8 M_\odot$, the PBH merger rate would be enhanced at $\sim t_c^\prime$ compared to the case without the effect of density perturbations. Accordingly, among the three spectra in fig. \ref{fig:Omega}, the blue one ($M=10^2 M_\odot$) peaks at $\sim\nu_2\sim 10^4 (M/M_\odot)^{-1}\ \text{Hz}=100\ \text{Hz}$; the orange one ($M=10^5 M_\odot$) peaks at $\sim\nu_2\sim 0.1\ \text{Hz}$ because $t_c^\prime \sim 4\times 10^{17}\ \text{s}$; and the red one ($M=10^8 M_\odot$) peaks at $z_\ast^{-1}\nu_2\sim 1000^{-1}\cdot 10^4 (M/M_\odot)^{-1}\ \text{Hz}=10^{-7}\ \text{Hz}$ with $t_\ast\sim 10^{13}\ \text{s}$ (the peak at a higher frequency $\sim 10^4 (M/M_\odot)^{-1}\ \text{Hz}=10^{-4}\ \text{Hz}$ comes from mergers near the present time).

\begin{figure}
\includegraphics[scale=0.33]{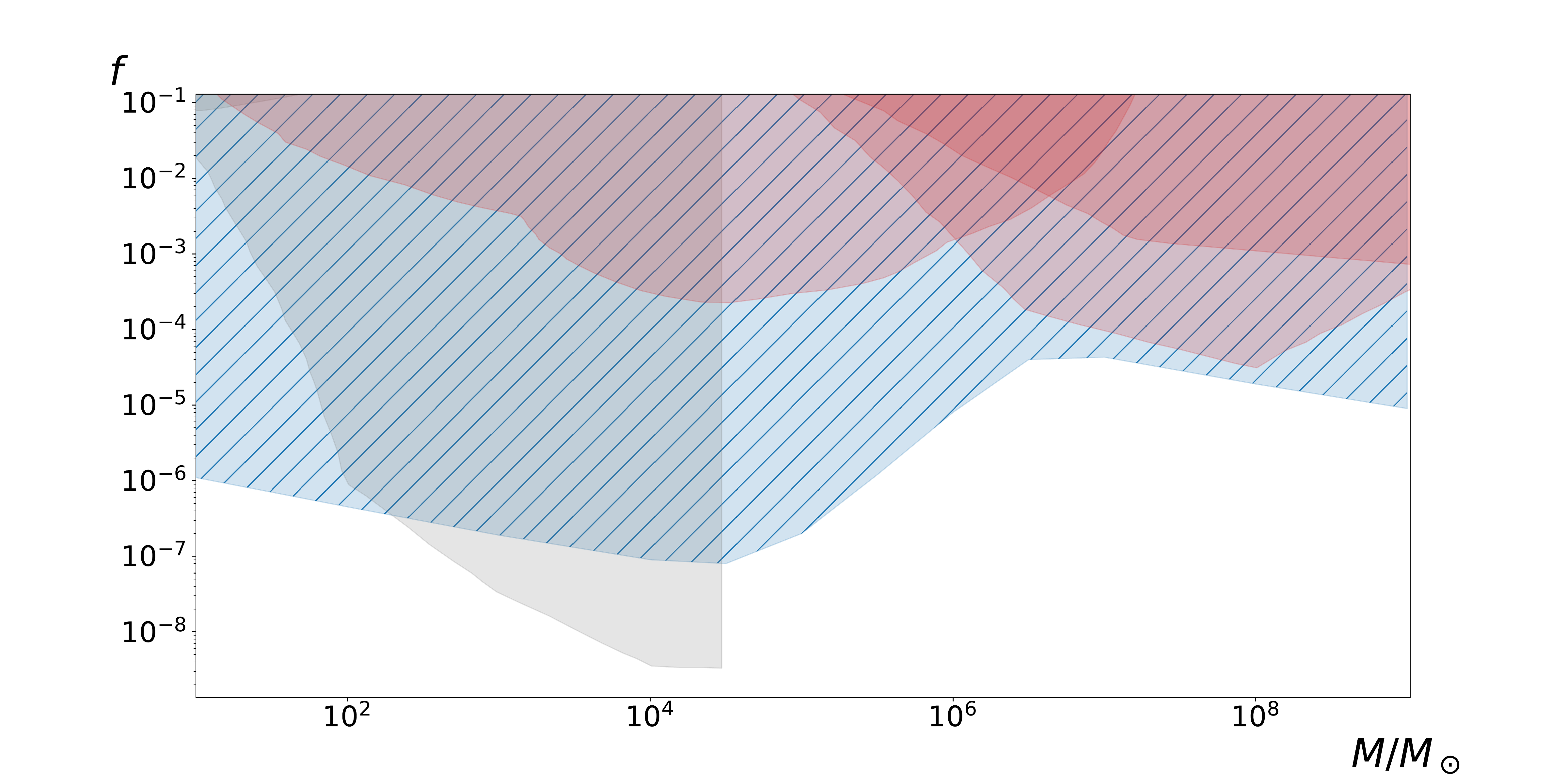}

\caption{\label{fig:f} Constraints on the fraction of dark matter in monochromatic
PBHs within the mass range $10\text{-}10^{9}M_{\odot}$. The gray and the red regions have been ruled out by observations. The blue shaded constraint is imposed by assuming that no GWB will be found by SKA or U-DECIGO.}
\end{figure}

The non-detection of GWB could thus place an upper
bound on the density of PBHs with the relevant masses. Similar
work was done in ref. \cite{Wang:2019kaf}, where the focus was on PBHs with mass
$M\lesssim M_{\odot}$. Here we are mainly interested in PBHs with $M\gtrsim 10M_\odot$. In fig. \ref{fig:f} we show the observational
constraints on PBHs within the mass range $10\text{-}10^{9}M_{\odot}$. The gray and the red regions (adapted from fig. 10 in ref. \cite{Carr:2020gox}) have been ruled out by observations: the red areas represent the non-observation of interactions between PBHs and other compact objects or cosmic structures \cite{Oguri:2017ock,Inoue:2017csr,Carr:2018rid}, and the gray represents the non-observation of the PBH accretion effect in CMB \cite{Serpico:2020ehh}.\footnote{Note that the gray and the red regions do not depend on the mechanisms of PBH formation. In the literature, the most stringent bound on large PBHs comes from the non-observation of the $\mu$-distortion in CMB generated by the dissipation of cosmological density perturbations, as investigated in refs. \cite{Carr:1993aq,Kohri:2014lza}. These studies basically rule out the formation of PBHs with $10^4 M_\odot \lesssim M \lesssim 10^{12}M_\odot$ in any appreciable numbers. This constraint is not shown in fig. \ref{fig:f}, because it was obtained by assuming Gaussian perturbations in the early universe, which is not a necessary condition for PBHs to form.} The non-detection of GWB can then be used  as an independent method to constrain PBHs: the blue shaded region could be excluded if SKA and U-DECIGO do not discover a GWB. As we can see, roughly speaking, these future missions could constrain the PBH abundance down to $f\lesssim 10^{-6}\text{-}10^{-4}$.

\section{Effect of mass accretion  \label{IV}}

An important aspect we did not take into account in the above discussion is the mass accretion of PBHs throughout the cosmic history. Although PBHs do not acquire much accretion when immersed in radiation fluid, they tend to absorb the ambient gas and dark matter after the time of dust-radiation equality and could grow by several orders of magnitude. However, a clear picture of how this happens is yet to be established due to the complex interactions between PBHs and the environment, especially when large scale structures begin to form at low redshifts \cite{Ricotti:2007jk,Ricotti:2007au,Chen:2016pud,Ali-Haimoud:2016mbv,Horowitz:2016lib,Ali-Haimoud:2017rtz,Poulin:2017bwe,Hutsi:2019hlw}. Therefore, most studies depend on some modeling parameters and assumptions. For this reason, in the following we shall simply discuss the effect of mass accretion on PBH mergers based on a toy model. 

Following refs. \cite{DeLuca:2020fpg,DeLuca:2020qqa}, we consider an accretion process that begins at redshift $z_i\approx 30$, before which the expansion rate of the universe is larger than the accretion rate; and stops instantaneously at a cut-off redshift $z_{\rm cut}=10$, which models the fact that structure formation and reionization may strongly suppress the accretion rate after $z_{\rm cut}$ \cite{Ricotti:2007au,Ali-Haimoud:2017rtz}. Let $t_{\rm cut}\sim 10^{16}\ \text{s}$ be the cosmic time at $z_{\rm cut}$. Since the coalescence time of a binary is dominated by the free-fall time only for mergers before $\sim 10^{13}\ \text{s}$, $t_{\rm cut}\gg 10^{13}\ \text{s}$ means the merger probability after $t_{\rm cut}$ is given by either case (a) or (b) in eq. (\ref{eq:P-4}). When mass accretion is not taken into account, $P(t)$ after $t_{\rm cut}$ can be re-expressed as\footnote{Here we have included the minimum value of $y$ in evaluating the integral (\ref{eq:P}) for case (a) (in contrast to eq. (\ref{eq:P-4})).}
\begin{equation}
P(t)\approx \frac{0.05f}{t}\begin{cases}
\left\{\left[1+\left(\frac{\sigma}{f}\right)^2\left(1-\left(\frac{t}{t_c^\prime}\right)^{-2/7}\right)\right]^{-29/37}\left(\frac{t}{t_c^\prime}\right)^{-1/7}\right.\\
\left.\ \ \ \ \ \ \ \ -f\left[1+\left(\frac{\sigma}{f}\right)^2\right]^{-21/74}\left(\frac{t}{t_{\rm last}}\right)^{3/8}\right\}, & t_c^\prime < t < t_{\rm last}, \\
\left(\frac{t}{t_c^\prime}\right)^{3/37}, & t < t_c^\prime,
\end{cases}\label{eq:P-2}
\end{equation}
where $t_{\rm last}=f^{-7}t_c\sim 10^{42}(M/M_\odot)^{-5/3}\ \text{s}$ is the time when the last merger event takes place: $P(t_{\rm last})\approx0$. Typically $t_{\rm last}$ is much larger than the current age of the universe. For later convenience we introduce
\begin{equation}
R(t)=P(t)t, \label{eq:RR}
\end{equation}
which can be interpreted as the probability (instead of the probability density) of merger events occurring within the time interval $(t,2t)$.

In the presence of accretion, the change in black hole mass  would certainly alter the binary orbit, and thus the PBH merger rate. By ref. \cite{DeLuca:2020qqa}, the evolution of the binary's semi-major axis follows $\dot{a}/a+3\dot{M}/M=0$, while the eccentricity remains constant. Then by the Peters formula (eq. (\ref{eq:merge1})), $t\propto a^4/M^3$, the coalescence time should decrease from $t$ to 
\begin{equation}
t_{\rm acc} = \left(\frac{M_f}{M_i}\right)^{-15}t,\label{eq:Racc}
\end{equation}
where $M_i$ is the initial PBH mass, and $M_f$ is the mass after accretion. The factor $\left(M_f/M_i\right)^{-15}$ can easily be a number $\ll \mathcal{O}(1)$, which means the coalescence time of binaries that had not merged before $z_i\approx 30$ could drop significantly after $z_i$, leading to a drastic change in the merger probability.

Let us consider PBHs with initial mass $M_i=100M_\odot$. The final mass $M_f$ is determined by the accretion rate, which is highly model dependent. Due to this uncertainty, we will take two different values of $M_f$. We first consider an ``extreme" case: $M_f=10^4M_\odot$. Without accretion, the last merger occurs at $t_{\rm last}\sim 10^{39}\ \text{s}$. After accretion, however, the coalescence time of this binary drops to $t_{\rm acc} \sim (10^4/10^2)^{-15}\cdot 10^{39}\ \text{s} = 10^9\ \text{s}$, which is much smaller than $t_{\rm cut}$. This means binaries that had not merged by $z_i$ would all merge by $z_{\rm cut}$!  

In fact, in this example, PBHs in binaries never get the chance to grow to $M_f$, because as their masses increase to a certain point, the coalescence time becomes comparable to $t_{\rm cut}$, then they would all merger at $z\sim z_{\rm cut}$. For instance, for a binary with coalescence time $t=t_{\rm last}\sim 10^{39}\ \text{s}$ in the case without accretion, when the PBH mass grows to $M$ such that
\begin{equation}
\left(\frac{M}{M_i}\right)^{-15}t_{\rm last} \sim t_{\rm cut},
\end{equation} 
which gives $M\sim 3000M_\odot < M_f=10^4M_\odot$, the merger occurs within a Hubble time at $t_{\rm cut}$. Since this event should belong to one of the last binaries, there would be no more merger events after $\sim t_{\rm cut}$.

The GWB spectrum could thus be rather different. In the case without accretion, a peak in the spectrum is from mergers near the present epoch, but for the above example, this peak would disappear, and a new peak would form at a lower frequency as a result of the aggregated merger events that were supposed to occur ``gradually" after $t_{\rm cut}$  (till $t_{\rm last}$). 

More precisely, to calculate the GWB spectrum, we need to add up signals from all mergers (of PBHs with mass $100M_\odot < M \lesssim 3000M_\odot$) that occur from $z_i$ to $z_{\rm cut}$. For simplicity, we consider discrete masses $M_k=10^{k/15}M_i$, where $k=(1,2,3,...,22)$, each one corresponding to an ``original" coalescence time $t\sim \left({M_k}/{M_i}\right)^{15}t_{\rm cut}=10^k t_{\rm cut}$. The GWB spectrum from these mergers is
\begin{equation}
\Omega_{\rm GW}^{\rm cut}(\nu_{\rm d})\sim \frac{n\nu_{\rm d}}{\rho_{c}z_{\rm eq}^{3}}\sum_{k=1}^{k_{\rm max}} R(10^k t_{\rm cut}) \left.\frac{\text{d}E_{\rm GW}(\nu_{\rm s}, M_k)}{\text{d}\nu_{\rm s}}\right|_{\nu_{\rm s}=\nu_{\rm d}(1+z_{k})},\label{eq:Omegacut}
\end{equation}
where $R(t)$ is given by eq. (\ref{eq:RR}), the maximum value of $k$ is given by $k_{\rm max}=\text{min}\left(\lfloor \log_{10}(t_{\rm last}/t_{\rm cut})\rfloor,\lfloor\log_{10}(M_f/M_i)^{15}\rfloor\right)$, and $z_k$ is the redshift when PBHs with mass $M_k$ merge ($\sim$ the redshift when PBHs grow to $M_k$). For simplicity, we assume the Bondi-type accretion rate $\dot{M}\propto M^2$, where the overdot stands for the derivative with respect to the cosmic time $t$. Noting that $M=M_i$ at $z_i$ and that $t\propto (1+z)^{-3/2}$ during the dust era, we obtain
\begin{equation}
\frac{\left(1+z_k\right)^{-3/2}-(1+z_i)^{-3/2}}{(1+z_{\rm cut})^{-3/2}-(1+z_i)^{-3/2}}=\frac{M_k^{-1}-M_i^{-1}}{M_f^{-1}-M_i^{-1}},
\end{equation}
which is a relation between $z_k$ an $M_k$ and can readily be inserted in eq. (\ref{eq:Omegacut}).

Let us now consider a less extreme example: $(M_i,M_f)=(100M_\odot,300M_\odot)$. In this case, the coalescence time of binaries with final mass $M_f$ is $t_{\rm acc}=\left(M_f/M_i\right)^{-15}t_{\rm last} > t_{\rm cut}$, which implies some binaries still remain after the aggregated mergers at $\sim z_{\rm cut}$. These binaries would merge after $t_{\rm cut}$ with probability density
\begin{equation}
P_{\rm acc}(t) = \left(\frac{M_f}{M_i}\right)^{15}P\left(\left(\frac{M_f}{M_i}\right)^{15}t\right), \label{P-acc}
\end{equation} 
where $P(t)$ is given by eq. (\ref{eq:P-2}). Depending on the values of $M_f$ and $t_{\rm last}$, mergers at this stage may stop before or after the present time.

In conclusion, in the presence of mass accretion, GWB comes from PBH mergers at three stages: (1) mergers before $z_i$; (2) aggregated mergers from $z_i$ to $z_{\rm cut}$; (3) (possible) mergers after $z_{\rm cut}$. The resulting GWB spectrum is
\begin{equation}
\begin{split}
\Omega_{\rm GW}^{\rm acc}(\nu_{\rm d})&\sim\frac{n\nu_{\rm d}}{\rho_{c}z_{\rm eq}^{3}}\left[\int^{z_{\rm max}}_{z_{i}} P(t)\frac{\text{d}t}{\text{d}z}\left.\frac{\text{d}E_{\rm GW}(\nu_{\rm s}, M_i)}{\text{d}\nu_{\rm s}}\right|_{\nu_{\rm s}=\nu_{\rm d}(1+z)}\text{d}z\right.\\&+\sum_{k=1}^{k_{\rm max}} R(10^k t_{\rm cut}) \left.\frac{\text{d}E_{\rm GW}(\nu_{\rm s}, M_k)}{\text{d}\nu_{\rm s}}\right|_{\nu_{\rm s}=\nu_{\rm d}(1+z_{k})}\\&+\left.\int^{z_{\rm cut}}_{z_{\rm min}} P_{\rm acc}(t)\frac{\text{d}t}{\text{d}z}\left.\frac{\text{d}E_{\rm GW}(\nu_{\rm s}, M_f)}{\text{d}\nu_{\rm s}}\right|_{\nu_{\rm s}=\nu_{\rm d}(1+z)}\text{d}z \right],\label{eq:Omegaacc}
\end{split}
\end{equation} 
where $z_{\rm min}$ is determined by whether the last merger occurs before or after today. If $(M_f/M_i)^{-15}t_{\rm last} \gtrsim 4\times 10^{17}\ \text{s}$, then $z_{\rm min}=0$. Otherwise $z_{\rm min}$ would be the redshift corresponding to the comic time $(M_f/M_i)^{-15}t_{\rm last}$.

\begin{figure}
\includegraphics[scale=0.33]{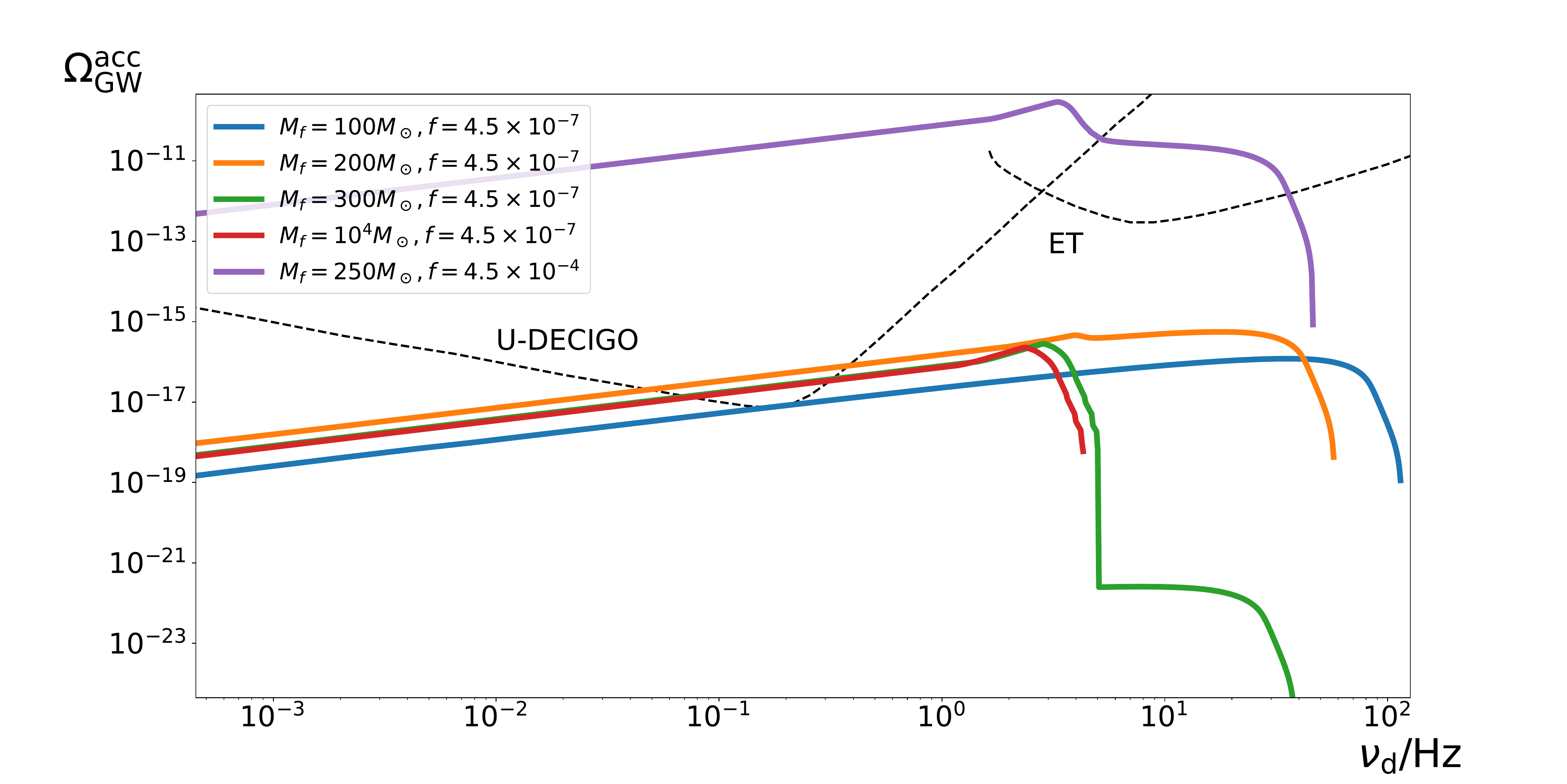}

\caption{\label{fig:Omega_acc} The five colored solid curves are examples of the energy
density spectrum of GWB from mergers of PBH binaries, with the effect of mass accretion included. Each spectrum
is determined by PBHs' initial mass ($M_i$), final mass ($M_f$), and the fraction of dark matter
in PBHs ($f$). These five examples have the same initial mass $M_i=100M_\odot$. The dashed curves are designed sensitivity curves of U-DECIGO and ET. We can clearly see that the spectra are rather sensitive to how much mass PBHs had attained during the accretion period.}
\end{figure}

In fig. \ref{fig:Omega_acc}, the red curve represents the spectrum $\Omega_{\rm GW}^{\rm acc}$ from $(M_i,M_f,f)=(100M_\odot,10^4M_\odot,4.5\times10^{-7})$. Compared with the blue curve, which is from the case of $M_i=10^2M_\odot$ without accretion, the red spectrum gets slightly enhanced at low frequencies from the aggregated mergers at $\sim z_{\rm cut}$, and gets suppressed at high frequencies because no merger events take place after $\sim z_{\rm cut}$. The green curve represents the spectrum from $(M_i,M_f,f)=(100M_\odot,300M_\odot,4.5\times10^{-7})$. Compared with the case of the red curve, there still are binaries merging after $\sim z_{\rm cut}$ till today following the probability $P_{\rm acc}(t)$, which contributes to the spectrum at $\nu_{\rm d}\lesssim 10^4 (M_f/M_\odot)^{-1}\ \text{Hz}\approx 30\ \text{Hz}$. 

By eqs. (\ref{eq:P-2}) and (\ref{eq:RR}), we have $R(t)\propto t^{3/37}$ for $t<t_c^\prime$, and $R(t)\propto t^{-1/7}$ for $t\gg t_c^\prime$. Then by eq. (\ref{P-acc}), $R_{\rm acc}(t)\equiv P_{\rm acc}(t)t$ scales as $R_{\rm acc}(t)\propto t^{3/37}$ for $t<t_{\rm acc}(t_c^\prime)$ and $R_{\rm acc}(t)\propto t^{-1/7}$ for $t\gg t_{\rm acc}(t_c^\prime)$. This means most mergers occur at around $t\sim t_{\rm acc}(t_c^\prime)$. Note that for $M_i=100M_\odot$, $t_c^\prime \sim 10^{26}(M_i/M_\odot)^{5/3}\approx 10^{22}\ \text{s}$. Hence, by eq. (\ref{eq:Racc}), most mergers during the accretion period involve PBHs with mass $M\sim (t_c^\prime/t_{\rm cut})^{1/15}M_i\sim (10^{22}/10^{16})^{1/15}\cdot 100M_\odot\approx 250M_\odot$. This explains why the peaks of the red and the green curves in fig. \ref{fig:Omega_acc} are both at $\nu_{\rm d}\sim z_{\rm cut}^{-1}\cdot 10^4(M/M_\odot)^{-1}\ \text{Hz}\approx 4\ \text{Hz}$.

For comparison, in fig. \ref{fig:Omega_acc} we also plot a spectrum with $(M_i,M_f,f)=(100M_\odot,200M_\odot,4.5\times 10^{-7})$ (orange curve). Compared with the case of the green curve, mergers after $\sim z_{\rm cut}$ brings a much larger GWB density at high frequencies. This can roughly be understood as follows.  For $M_f=300M_\odot$, $t_{\rm acc}(t_c^\prime)\sim 10^{15}\ \text{s}$. Then by eq. (\ref{eq:P-2}), the merger probability after $t_{\rm cut}\sim 10^{16}\ \text{s}\gg t_{\rm acc}(t_c^\prime)$ is
\begin{equation}
R_{\rm acc}(M_f,t)\sim 0.05f \left(\frac{\sigma}{f}\right)^{-58/37}\left(\frac{t}{10^{15}\ \text{s}}\right)^{-1/7}.
\end{equation}
For $M_f=200M_\odot$, $t_{\rm acc}(t_c^\prime)\sim 10^{17}\ \text{s}$, when the merger probability takes the maximum value $R_{\rm acc}(M_f,10^{17}\ \text{s})\sim 0.05f$. Therefore, the ratio of the number of merger events for $M_f=200M_\odot$ to that for $M_f=300M_\odot$ from $t_{\rm cut}$ to the present time can be estimated as
\begin{equation}
\frac{R_{\rm acc}(200M_\odot, 10^{16}\ \text{s})}{R_{\rm acc}(300M_\odot, 10^{17}\ \text{s})}\sim 10^6.
\end{equation}
This explains the huge difference between the two spectra (green and orange curves) at high frequencies in fig. \ref{fig:Omega_acc}.

For a lager value of $f$, features from the effect of accretion at the high-frequency end could be accessible to potential probes such as ET and CE. An example (purple curve) is shown in fig. \ref{fig:Omega_acc}.

Generally speaking, including the effect of mass accretion suppresses the GWB density at high frequencies, but the low-frequency end is not significantly influenced (less than an order of magnitude). In the above examples, GWB spectra from cases with $M_i=10^2M_\odot$ intersect the U-DECIGO curve at the low-frequency end, so the possible constraint on the PBH abundance is still $f\sim 10^{-6}$.  The exact form of a spectrum given $M_i$ is obviously rather sensitive to the accretion process, a clear description of which is still lacking. Hence, in this work, we shall not further discuss the impact of accretion on the PBH bound.

\section{Conclusions and discussion  \label{V}}

In this work, we have revisited the merger rate of large PBHs, noticing that the Peters formula, which is often adopted in the literature as the coalescence time of PBH binaries, is invalid if the orbital period of the binary is larger than what the formula gives. Applying the Peters formula could underestimate the merger rate of PBH binaries merging before the time $t\sim 10^{13}\ \text{s}$.

We further estimated the energy density spectrum of GWB from PBH mergers, and found that merger events at $t\sim 10^{13}\ \text{s}$ could bring an extra peak to the spectrum for PBHs with mass $M\gtrsim 10^8 M_\odot$. If PBHs have mass within the range $10^5\text{-}10^8 M_\odot$, the peak would be dominated by mergers occurring at $t\sim 10^{26}(M/M_\odot)^{-5/3}\ \text{s}$. For a sufficiently large PBH abundance, the resulting GWB spectrum could possibly be within the reach of potential GW probes such as SKA and U-DECIGO. If future missions do not see such a background, the fraction of dark matter in PBHs is constrained to $f\lesssim 10^{-6}\text{-}10^{-4}$ within the mass range $10\text{-}10^9M_\odot$.

Mass accretion on black holes is inevitable at low redshifts, and is especially crucial for the evolution of large PBHs. By considering a simple model, where most accretion occurs between redshift $z\approx 30$ and $z\approx 10$, we found that the GWB spectrum at the high-frequency end is sensitive to how much mass PBHs had attained. Even if there was only a small growth in the black hole mass, PBH binaries that had not merged before $z\approx 30$ could all merge before $z\approx 10$ because the coalescence time could drop significantly. Accordingly, the GWB density gets suppressed at high frequencies. For PBHs with initial mass $M_i=\mathcal{O}(10\text{-}100)M_\odot$, the GWB spectrum with features affected by accretion could be captured by potential GW probes such as ET and CE if the PBH abundance is sufficiently large.

In the presence of mass accretion, fig. \ref{fig:f} needs modifications and further interpretation \cite{DeLuca:2020fpg}. Note that the gray region in this figure was obtained by analyzing the accretion effect in CMB (at $z\lesssim 1000$), while the red regions were found by considering the possible interactions between PBHs and galaxies or other compact objects (at low redshifts). Strictly speaking, these two constraints should not be placed in one figure considering that PBHs with $M_i \gtrsim 10M_\odot$ could grow by orders of magnitude from high to low redshifts \cite{DeLuca:2020fpg}. Let $f_i$ ($f_f$) be the initial (final) fraction of dark matter in PBHs. If the horizontal axis and the vertical axis in fig. \ref{fig:f} are respectively regarded as $M_f$ and $f_f$, one ought to distort the orange and the gray regions. Since the comoving  number density of PBHs should remain the same over time, we have $f_f\sim(M_f/M_i)f_i$. For want of a specific accretion model, we do not intend to go further in this work, but apparently the constraints are expected to be weaker.  

Lastly, we note that the condition for mergers at $z\sim 1000$ to play an important role in the GWB spectrum does not apply to $(M,f)=(30M_\odot, 10^{-3})$, which could be an explanation of the LIGO-Virgo events. Therefore, GWB from the mergers of these possible PBHs found in, e.g., refs. \cite{Wang:2016ana,Raidal:2017mfl} is not influenced by the ``new" term $t_{\rm ff}$ in the coalescence time. However, the effect of mass accretion discussed in section \ref{IV} could be relevant, depending on how much PBHs had grown before their mass reached $30M_\odot$. The resulting GWB should be modified accordingly. We leave the analysis of this problem for future work.

\section*{Acknowledgments}
I am grateful to Tanmay Vachaspati for insightful comments on the manuscript. This work is supported by the U.S. Department of Energy, Office of High Energy Physics, under Award
No. de-sc0019470 at Arizona State University.

\bibliography{Peterslib}
\end{document}